\documentstyle[preprint,prd,aps,floats,epsfig]{revtex}
\tighten
\begin{document} 
\draft
\preprint{
\begin{tabular}{rr}
CfPA/96-th-21\\
IMPERIAL/TP/96-97/01 \\
DAMTP-96-84\\
Submitted to  PRD
\end{tabular}
}

\title{Non-Gaussian Spectra}
\author{Pedro G. Ferreira$^{1}$ and Jo\~ao Magueijo$^2$}
\address{$^{(1)}$Center for Particle Astrophysics,
University of California, Berkeley  CA 94720-7304,USA\\
$^{(2)}$The Blackett Laboratory, Imperial College,
Prince Consort Road, London SW7 2BZ, UK}
\maketitle
\begin{abstract}
Gaussian cosmic microwave background skies are fully specified
by the power spectrum. The conventional method of 
characterizing non-Gaussian skies is to evaluate higher order
moments, the n-point functions and their Fourier transforms.
We argue that this method is inefficient, due to the redundancy of
information existing in the complete set of moments.
In this paper we
propose a set of new statistics or non-Gaussian spectra
to be extracted out of the angular distribution of the Fourier
transform of the temperature anisotropies in the
small field limit.
These statistics complement the power spectrum and act as
localization, shape, and connectedness statistics.
They quantify generic non-Gaussian structure, and may
be used in more general image processing tasks. 
We concentrate on a subset of these statistics and argue that while they 
carry no information in 
Gaussian theories they may be the best arena for making predictions
in some non-Gaussian theories. 
As  examples of applications we consider superposed Gaussian 
and non-Gaussian signals,
such as point sources in Gaussian theories or the realistic
Kaiser-Stebbins effect.  We show that in these theories 
non-Gaussianity is only
present in a ring in Fourier space, which is best isolated in our
formalism.  Subtle but strongly non-Gaussian theories are
also written down for which only non-Gaussian spectra may accuse 
non-Gaussianity.
\end{abstract}

\date{\today}


\renewcommand{\thefootnote}{\arabic{footnote}}
\setcounter{footnote}{0}
\section{Introduction}
Gaussianity plays a central role in current
theories of structure formation  \cite{Peebles2}. 
Inflationary theories are normally
invoked to justify Gaussianity \cite{inflCMBR}  but, historically, 
simplicity was perhaps what first motivated this assumption. 
As data has started to flood cosmology, however, the problem  of 
testing Gaussianity has reappeared both in Cosmic Microwave Background
(CMB) analysis \cite{kogut}, 
and galaxy survey analysis \cite{lascampanas}. A trend in data
analysis has been established which relies on Gaussianity and a
lingering feeling exists that the whole thing might fall through
should the data prove to be non-Gaussian in the first place. Furthermore 
structure formation theories exist which in one way or another predict 
non-Gaussian primordial fluctuations. Cosmic strings  and 
textures \cite{VS} provide two such examples. Pinning down
what precise non-Gaussian predictions such theories can make
is a task crying for a comprehensive formalism for quantifying
general non-Gaussianity. Finally even if the ``signal'' is Gaussian 
it may happen that a non-Gaussian noise component is present, eg. 
unresolved point sources \cite{catpeople}. 
A precise prediction of their observational
properties could then assist in their subtraction from data,
before the final theoretical analysis is performed. 

One is therefore left with the problem of how to test Gaussianity,
and how to quantitatively specify the most general non-Gaussian theory.
Several tests for non-Gaussianity have been proposed in the past.
Peaks' statistics \cite{BE,Jusk}, topological tests \cite{Pcoles,gott}, 
the 3-point correlation function \cite{kogut,gangui}, 
skewness and kurtosis \cite{scaramelo,periv}, and temperature and temperature
gradient histograms \cite{coulson} are the most topical examples. In some
cases these tests were only shown to be applicable for rather
artificial non-Gaussian distributions \cite{Pcoles}. In other cases the
tests were applied only to extremely non-Gaussian signals,
or the eroding effects of Gaussian noise were not explored \cite{pst}.

These tests however,  are by no means exhaustive.
One can always devise a non-Gaussian theory which evades detection
by any of these tests, even when the hard realities of experiment
do not fully erase signal non-Gaussianity.
The only way to fully ascertain Gaussianity is to apply to data
a comprehensive formalism for encoding non-Gaussianity in its broadest
generality.
The $n$-point correlation function provides such a framework,
and it has long been used in cosmology \cite{Peebles1} and other branches of
physics \cite{randfields}. Computing the $n$-point function for large $n$
is however a practical impossibility. Taking the COBE data as an example
\cite{kogut}, only the 3-point function has been computed, and even in that
case attention was restricted to the pseudo-collapsed 
and equilateral slices. 

In section 2 we start off by showing how the $n$-point correlation 
functions for $n$ up to any $N>3$ contain redundant information.
For Gaussian fields all the $N>2$ correlators can be determined
from the 2-point correlator. We show that even for the most general 
non-Gaussian theory information encoded in the
$N>3$ correlator is dependent on information in lower order correlators.
Furthermore we show that one can never be sure  that
by truncating the infinite correlator series at some $N$ one has 
all the information about the most general non-Gaussian theory. 
Strongly non-Gaussian theories may be written down which 
have Gaussian moments up to any given order $N$.
The $n$-point function formalism then appears to have two drawbacks: 
redundancy and impractical complexity. We shall argue that these two 
drawbacks are due to each other, and that they may be eliminated 
altogether. 

In this paper, in Section~\ref{newstats}, we propose an alternative 
formalism for comprehensively encoding non-Gaussianity. In the guise
to be used in this paper the formalism lives naturally
in Fourier space, and we have chosen to highlight non-Gaussianity
other than in the phases.
The idea of looking for non-Gaussianity in Fourier space has been
disfavoured in the past. It is argued that localized non-Gaussianity
in real space
(such as what is produced in cosmic string or textures scenarios)
 will be obscured in Fourier space
due to the central limit theorem. It is also often assumed that
a Gaussian field can be accurately modelled as the Fourier transform
of a field whose randomness is solely in the phases. However, as we
argue in section 2, looking in Fourier space allows us to probe
the non-Gaussian nature of the field at specific scales, a fact which
is particularly useful when one can model the field as combination
of a Gaussian field which dominates on certain scales and a
non-Gaussian field which dominates on others.
Another very strong reason for considering Fourier space statistics
seriously is the fact that the highest resolution measurements
of CMB anisotropies will be performed by interferometric devices,
which naturally measure quantities in Fourier space (the ``$uv$ plane'').

Therefore, ignoring prejudice,
in section 3 we define a set of ``non-Gaussian spectra'' in terms of the
Fourier transform of the temperature anisotropies. Our
definitions follow up the proposals in \cite{mag1}, but
they are substantially more practical. We then characterize
the probability distribution function of these spectra in Gaussian theories
and in Appendix II give a physical interpretation of the qualities which 
they  measure. We set these quantities up so that while
they contain all the information degrees of freedom, they
do away with any redundancy. As a result we come up with a formalism which
shares with the $n$-point correlators the property of being comprehensive,
but with the advantage that it is computable and non-redundant.
Within the large set of statistics considered in this paper we 
concentrate on a set of statistics which only use the information in
the absolute value of the Fourier modes. These are grouped in
two types of spectra: the ring spectrum and the inter-ring spectrum. 
For the sake of maximal originality we leave to a future  publication the
investigation of the role played by the more prosaic phase information.

In section 4 we consider three different applications. Firstly
we consider the case of a point source which is obscured
by Gaussian fluctuations. Secondly we consider the realistic
temperature anisotropy induce by a cosmic string, including
both the post-recombination Kaiser-Stebbins effect and the
Gaussian fluctuations at the surface of last scattering.
Finally we construct a strongly non-Gaussian theory, a theory
which produces skies which have a zero probability of 
occurring in a Gaussian theory. To all these examples we apply
a battery of conventional statistics and show that they
evade any detection of non-Gaussianity. We show however that our
statistics reveal the non-Gaussian nature of the skies.

In section 5 we conclude by discussing the limitations of
these statistics and possible extensions.

\section{The $n$-point correlation function}
We start by reviewing the $n$-point correlation function formalism.
We then introduce the concept of $uv$-plane invariants, that is
quantities which are made up of Fourier modes $a({\bf k})$, 
and which are invariant under rotations and translations.
We show how the Fourier transform of the $n$-point correlation
function is made up of $uv$-plane multilinear invariants.
One may then count the number of degrees of freedom in
the Fourier modes for a given sky coverage. By doing so we
show that the  $n$-point correlators for $n$ up to a certain
$N$ contains information which can only be redundant. 
This will set the tone for the next Section: trying to do 
away with the redundancy and complexity of the $n$-point correlation
function.
 
\subsection{The $n$-point correlation function and its transform}
We consider CMB data in the small angle limit, when projecting onto
a planar patch is suitable. Since data may come in either real or
Fourier space we hope to address the problem of non-Gaussianity
in terms of these two descriptions. In this paper however
we will concentrate on the Fourier space description, and thus
produce statistics better suited to interferometers.
We shall use the convention:
\begin{equation}\label{fourier}
  \frac{\Delta T({\bf x})}{T}
={\int {d{\bf k}\over 2\pi}a({\bf k})e^{i{\bf k}\cdot{\bf x}}}
\end{equation} 
The $n$-point correlation function is defined as the expectation
value of the product of any $n$ temperatures. Translational and
rotational invariance make redundant the position of one of the points
and the direction of another.
Hence the $n$-point function may be written as a function of
$(x_2, {\bf x}_3,\cdots,{\bf x}_n)$ in the form
\begin{equation}
  C^n(x_2, {\bf x}_3,\cdots,{\bf x}_n)={\langle 
  \frac{\Delta T({\bf x}_1)}{T}...\frac{\Delta T({\bf x}_n)}{T} 
  \rangle}
\end{equation}
The 2-point correlation function and its Fourier transform, the 
angular power spectrum $C(k)$,  are well-known.
They fully specify Gaussian fluctuations. For Gaussian fluctuations
non-vanishing higher order correlation functions exist, 
but they are redundant as they can be obtained from 
the two-point correlation function.
This is not the case in non-Gaussian theories, for which the $n$-point
correlators act not only as a non-Gaussianity indicator, 
but are also an indispensable fluctuation qualifier, as much as the 
power spectrum. 

The angular power spectrum may be generalized for $n>2$ by
Fourier analysing the $n$-point  function 
\begin{eqnarray}\label{cn1}
  &C^n&(x_2, {\bf x}_3,\cdots,{\bf x}_n)= \nonumber \\
&&\int \frac{dk_2}{(2\pi)^{1/2}}
  \cdots \frac{d{\bf k}_n}{2\pi}C^n(k_2, \cdots,{\bf k}_n)
  {e^{ik_2 x_2} \cdots e^{i{\bf k}_n\cdot{\bf x}_n}}
\end{eqnarray}
In general $C(k)$ is more predictive than $C(x)$, as it tells us how much
power exists on a given scale. In the same way one may expect
the transform $C^n(k_2, \cdots,{\bf k}_n)$ to be more predictive
than its configuration space counterpart, as it
tells us how much non-Gaussianity exists on each scale.
We shall call $C^n(k_2, \cdots,{\bf k}_n)$ a non-Gaussian
spectrum. One may also define Gaussian spectra as  correlators
of the $a({\bf k})$ modes:
\begin{equation}\label{cn2}
  {\langle a({\bf k}_1)\cdots a({\bf k_n})\rangle}=
  \delta({\bf k}_1+\cdots {\bf k_n})C^n(k_2, \cdots,{\bf k}_n)
\end{equation}
where the $\delta$ function and functional form of $C^n$ result
from the requirements of translational and rotational invariance
(see (\ref{rott}) below). Using (\ref{fourier}) one may easily 
check that the two definitions (\ref{cn1}) and (\ref{cn2})
of $C^n(k_2, \cdots,{\bf k}_n)$ agree.

Non-Gaussian spectra are more
complicated than power spectra, since they are functions
of many variables. As $n$ increases one is left with the problem
of how to pack so much information. We will however show that most
of the information encoded in $C^n(k_2, \cdots,{\bf k}_n)$ is largely
redundant, even for the most general non-Gaussian fluctuation.

\subsection{$uv$-plane multilinear invariants as components of
the $n$-point correlation function}
Here we show an equivalent route to non-Gaussian spectra.
This route draws on work in \cite{mag1}, where the spherical harmonic
coefficients $a^\ell_m$ are used to define quantities other than 
the $C_\ell$ spectrum which are invariant under the 3D rotation group.
$m$-spectra and inter-$\ell$ correlators appear as supplementary
information. These spectra are multilinear combinations of
the $a^\ell_m$ which can be generally written as sums of products
of Clebsch-Gordon coefficients. It can be shown that they act
as a decomposition of the $n$-point function on the sphere in 
a suitable base made up of Legendre polynomials and spherical harmonics.
These spectra are trivial to implement on 
a computer, but are formally quite complicated for large $\ell$. 
An explicit  expression for the quadrupole $m$-shape was given in 
\cite{mag1} with a suggested application to texture scenarios. 
Fortunately at very high $\ell$ one may simply reformulate the
problem in terms of  the Fourier representation of small
patches.  $m$-spectra and inter-$\ell$ correlators then become very simple. 
They reappear as $uv$-plane invariants, that is quantities made
up of the $a({\bf k})$ modes, and which are invariant under
2D rotations and translations (the projected 3D rotation group).

The non-Gaussian spectra $C^n(k_2, \cdots,{\bf k}_n)$  are
invariant under rotations and translations. This requirement
may also be imposed on any set of qualifiers of a random field
which statistically satisfies these invariances.
Under a rotation $R_\theta$ and a translation along a vector 
${\bf t}$ the Fourier  components transform as in
\begin{eqnarray}\label{rott}
  R_\theta(a({\bf k}))&=&a(R_\theta({\bf k}))\nonumber\\
  T_{\bf t}a({\bf k})&=&e^{i{\bf k}\cdot {\bf t}}a({\bf k})
\end{eqnarray}
A systematic way to generate invariants out of the $a({\bf k})$
is to consider multilinear combinations, that is sums of products
of $n$ modes $a({\bf k)}$ (monomials). For these to be invariant
under translations it is necessary that the vectors ${\bf k}_i$
used in each monomial add up to zero. To achieve invariance under 
rotations one must then, for each monomial, average over all possible 
rotations of the ${\bf k}_i$ configuration used.
One may formally write the most general multilinear invariant 
of order $n$ as
\begin{equation}\label{geninv}
  I^{(n)}={1\over N_\theta}\sum_{\theta}\prod_{i=1}^n a({\bf k}_i)
\end{equation}
in which the vectors ${\bf k}_i$ considered in each product must
add up to zero and always take the same configuration, and $N_\theta$ is the
total number of possible rotations of the configuration, should
Fourier space be discretized.
For $n=2$ the only invariant for each $k$ 
is the angular power spectrum. Given
a vector ${\bf k}$, the requirement that the second vector 
in the binomial adds to zero fully determines the second vector.
Averaging over all rotations makes the direction of the 
first vector irrelevant. The invariant (\ref{geninv}) then 
reduces to 
\begin{equation}
  I^{(2)}(k)={1\over N}\sum_{|{\bf k}|=k} |a({\bf k})|^2
\end{equation}
For the third order invariants one now has an invariant which depends
on a vector and a scalar. Independent invariants are parameterized
by the third vector and  the relative direction of the second vector.
The first vector is fully determined by the requirement that the 3 vectors
add to zero. The actual directions of the second and third vectors 
are made redundant by taking the circular 
average. A particularly interesting 3rd order invariant
may be obtained if one demands that the 3 vectors
used all have the same moduli. Then for each $k$ only one invariant
exists, the one obtained with the configuration plotted in Figure 1.

Diagrammatically one may then write down the most general invariant
for any order, rapidly bumping into unwanted proliferation.
The procedure however is very simple, and reduces to Eqn.~(\ref{geninv})
and the various independent diagrams it allows. 
The most general multinomial invariant of degree $n$ is 
a function of $(k_2,{\bf k}_3, \cdots, {\bf k}_n)$.
Hence the non-Gaussian spectra defined in the decomposition 
of the $n$-point correlation function correspond to the most general
multinomial invariant one may construct out of the $a({\bf k})$.

\subsection{Exposing the redundancy of the $n$-point function}
The approach just devised has the advantage of allowing us to 
expose the redundancy of the $n$-point correlation function.
Let us start by counting the number of degrees of freedom
present in the Fourier modes produced by a given measurement. 
If we had full sky coverage then there would be
$2k+1$ modes per unit of $k$. Finite sky coverage has the effect
of correlating neighbouring modes among these,
thereby reducing the number of independent modes per unit 
of $k$ to $2kf_{\rm sky}$, if $f_{\rm sky}$ is the fraction of 
sky covered. An alternative Fourier space
discretization is then required, so that the modes in the new
mesh are quasi-uncorrelated, while encoding all the statistical
information in the original modes. This may be done with a
so-called uncorrelated mesh (see \cite{hobsmag}). There is
some arbitrariness in where the new mesh in laid. This arbitrariness allows
us to be sloppy with the invariances imposed in the previous section,
since any vector ${\bf k}$ may now be placed anywhere in the 
uncorrelated mesh cell. Hence the angles required by configurations
such as the ones in Fig. 1 should be seen as flexible, as
far as the mesh resolution in concerned.

Let us now consider a generic $\Delta k=1$ ring containing 
$N_{ring}\approx 2kf_{\rm sky}$ uncorrelated mesh points.
Since there are 3 degrees of freedom in rotations and translations
one may not build more than $N_{ring}-3$ independent invariants 
per unit of $k$, plus 3 invariants relating adjacent 
$\Delta k=1$ rings. The number of multilinear invariants making up 
the $n$-point function transform is vastly larger. Even 
if we restrict ourselves to invariants made up only of
modes in each ring, the number of invariants is 1 for $n=2,3$
(see Figure 1),
then, for $n>3$, of order ${\cal O}(N_{ring}^{n-3})$, if $N_{ring}\gg 1$.

The situation gets worse if we consider inter-ring multinomial
invariants. Let us now consider a square in Fourier space with
$N_p\times N_p$ uncorrelated mesh points.
Then for large $N_p$ the number of multilinear invariants
of order $n$ in all rings is of order ${\cal O}(N_p^{n-1})$.
The number of independent mesh points, on the contrary, is
of order ${\cal O}(N_p^2)$.

Hence there must be an algebraic dependence between all the
multilinear invariants. The information encoded in the higher
order correlators must therefore repeat itself in any theory, Gaussian
or not. We therefore argue that the $n$-point function formalism,
while comprehensive, is not systematic. This is not to say that
some truncation of the correlator series might not be useful
as a non-Gaussianity test. In particular we feel that ring
multinomial invariants, such as the cubic one depicted in Figure 1,
may be useful non-Gaussianity tests.
 

\section{Ring and inter-ring spectra}\label{newstats}
We now propose an alternative packaging for the information in Fourier
space. Comparing it with the $n$-point transform, it is simpler, 
does away with redundancy, and has an immediate physical interpretation.
We divide the $uv$-plane in $\Delta k=1$ rings where 
$N_k=2kf_{\rm sky}$ independent modes lie.  Out of these we may
build $N_k-3$ invariants. In whatever we do we shall always make
sure that the formalism proposed produces the power spectrum $C(k)$
as the first of these quantities.
The other $C(k,m)$, for $m=1,\cdots, N_k-4$, are the ring spectra.
We shall not consider multilinear invariants, but shall search for
alternative prescriptions.  On top of these, for each two adjacent
rings there will be 3 invariants, the inter-ring correlators.
Given the arbitrariness of the Fourier
modes mesh exact position we may also be justified in building
simply $N_k$ non-invariant quantities for each ring, as long 
as we know how they transform. We found the latter attitude 
more practical, but shall give in Appendix I the correct prescription
for building properly invariant quantities.

For a Gaussian theory the probability of a given map depends
only on the map power spectrum. Consider then a very non-Gaussian map
by which we mean something we visually recognize as very structured.
Consider also various other maps with the same power spectrum,
but which we visually recognize as very Gaussian.
All these maps, Gaussian looking or not, have the same probability 
in Gaussian theories. In non-Gaussian theories, on the contrary,
the probability of a given map depends on more than its power spectrum. 
Hence, within the set of maps considered above, it may happen
that the non-Gaussian looking map is now considerably more
probable than the other maps.
The point we wish to make is that non-Gaussianity arises not from structured
maps being less likely in Gaussian theories, but from structured maps being
more likely in non-Gaussian theories.

This seemingly innocent remark has two important implications.
First it implies that the natural variables for non-Gaussianity
spectra should  be uniformly distributed in Gaussian theories.
In contrast, at least in some non-Gaussian theories, the same
variables should have peaked distributions. 
Hence non-Gaussian spectra should carry no information 
whatsoever in Gaussian theories, but they should be  
highly predictive at least in some non-Gaussian theories.

A second implication is that disproving Gaussianity on its
own merits is a contradiction in terms. One can always disprove a 
given non-Gaussian theory on its own merits 
by measuring a non-Gaussian spectrum 
and finding it to be away from the theoretically predicted ridge.  
However, any non-Gaussian spectrum measurement is
equally probable in Gaussian theories, and so it can never be used as 
evidence against Gaussianity. Disproving Gaussianity is
then a matter dependent on the available competing non-Gaussian theories. 
If one measures a non-Gaussian spectrum spot on the prediction
of a well motivated non-Gaussian theory then this is strong evidence 
in favour of that non-Gaussian theory. One may simply argue that 
the non-Gaussian theory has predicted the observation with 
much larger probability than the Gaussian theory. Pedantically,
the observation has not disproved Gaussianity. However it
has discredited Gaussianity  massively in the face of the more predictive
competing non-Gaussian theory.

It is under the  requirement that non-Gaussian spectra ought to be 
uniformly distributed in Gaussian theories that we now proceed to define
non -Gaussian spectra.
Consider a ring of the $uv$-plane where $N_k$ independent
complex modes $a({\bf k}_i)=\Re [a({\bf k}_i)] 
+i\Im [ a({\bf k}_i)]$ live.
In Gaussian theories these are distributed as
\begin{eqnarray}
  F(\Re[a({\bf k}_i)],\Im[a({\bf k}_i)])=
      {1\over (2\pi\sigma^2)^{N_k/2}}\times\nonumber \\
        \exp{-\left(
          {1\over 2\sigma_k^2}\sum_{i=1}^{m_k}(\Re^2[a({\bf k}_i)]+
          \Im^2[a({\bf k}_i)])\right)}
\end{eqnarray}
where $m_k=N_k/2$.
First separate the $N_k$ complex modes into $m_k$ moduli $\rho_i$
and $m_k$ phases $\phi_i$
\begin{eqnarray}
  \Re[a({\bf k}_i)]&=&\rho_i\cos{\phi_i}\nonumber\\
  \Im[a({\bf k}_i)]&=&\rho_i\sin{\phi_i}
\end{eqnarray}
The Jacobian of this transformation is 
\begin{equation}\label{jac1}
  \left| {\partial (\Re[a({\bf k}_i)],\Im[a({\bf k}_i)])
      \over \partial(\rho_i,\phi_i)}\right|=\prod_{i=1}^{m_k}\rho_i
\end{equation}
The  $\{\rho_i\}$ may  be seen as 
Cartesian coordinates which we transform into polar coordinates.
These consist of a radius $r$ plus $m_k-1$ angles $\tilde\theta_i$
given by
\begin{equation}\label{pol}
  \rho_i=r\cos{\tilde\theta_i}\prod_{j=0}^{i-1}\sin{\tilde\theta_j}
\end{equation}
with $\sin{\tilde\theta_0}=\cos{\tilde\theta_{m_k}}=1$.
In terms of these variables the radius is related to the
angular power spectrum by $C(k)=r^2/(2m_k)$. In general the first 
$m_k-2$ angles $\tilde\theta_i$ vary between $0$ and $\pi$ and the 
last angle varies between 0 and $2\pi$.
However because all $\rho_i$ are positive all angles are in $(0,\pi/2)$.
The Jacobian of this transformation is 
\begin{equation}\label{jac2}
  \left| {\partial(\rho_1,\cdots,\rho_{m_k})
      \over \partial(r,\tilde\theta_1,\cdots,\tilde\theta_{m_k-1}}\right|=
     r^{m_k-1}\prod_{i=2}^{m_k-1}\sin^{m_k-i}{\tilde\theta_{i-1}}
\end{equation}
Polar coordinates in $m_k$ dimensions may be understood as the iteration
of the following rule:
\begin{eqnarray}
  \rho_i&=&r_i\cos{\tilde\theta_i}\nonumber \\
  r_{i-1}&=&r_i\sin {\tilde\theta_i}
\end{eqnarray}
in which $r_i$ is the radius of the shade $m_k-i+1$ dimensional sphere
obtained by keeping fixed all $\rho_j$ for $j=1,\cdots,i-1$:
\begin{equation}
  r_i={\sqrt{\rho_i^2+\rho_{i+1}^2+\cdots+\rho_{m_k}^2}}
\end{equation}
One may easily see that this is how 3D polars work, and also that
the transform (\ref{pol}) follows this rule.
Hence one may invert the transform (\ref{pol}) with
\begin{equation}
  \tilde\theta_i=\arccos{\rho_i\over {\sqrt{\rho_i^2+\rho_{i+1}^2+
        \cdots+\rho_{m_k}^2}}}
\end{equation}
for $i=1,\cdots,m_k-1$. 

The total Jacobian of the transformation
from $(\Re[a({\bf k}_i)],\Im[a({\bf k}_i)])$ 
to $\{r,\tilde\theta_i,\phi_i\}$ is
just the product of (\ref{jac1}) and (\ref{jac2}).
Hence for a Gaussian theory one has the distribution
\begin{equation}
  F(r,\tilde\theta_i,\phi_i)={
    {r^{N_k-1}\exp{-\left(r^2\over 2\sigma_k^2\right)}}
      \over (2\pi\sigma^2)^{N_k/ 2}}
          \prod_{i=1}^{m_k-1}\cos{\tilde\theta_i}
          (\sin{\tilde\theta_i})^{N_k-2i-1}
\end{equation}
In order to define $\tilde\theta_i$ variables 
which are uniformly distributed in 
Gaussian theories one may finally perform the transformation on each
$\tilde\theta_i$:
\begin{equation}
  {\theta_i}=\sin^{N_k-2i}(\tilde\theta_i)
\end{equation}
so that for Gaussian theories one has:
\begin{equation}
  F(r,{\theta}_i,\phi_i)={r^{N_k-1}e^{-r^2/(2\sigma_k^2)}
    \over 2^{m_k-1}(m_k-1)!}\times 1 \times\prod_{i=1}^{m_k}{1\over
    2\pi}
\end{equation}
The factorization chosen shows that all new variables are independent
random variables for Gaussian theories. $r$ has a $\chi^2_{N_k}$
distribution,
the ``shape'' variables $\theta_i$ are uniformly distributed
in $(0,1)$, and the phases $\phi_i$ are uniformly distributed in $(0,2\pi)$.

The  variables $\theta_i$ define a non-Gaussian shape spectrum,
the {\it ring spectrum}. 
They may be computed from ring moduli $\rho_i$  simply by
\begin{equation}
  {\theta}_i={\left(\rho_{i+1}^2+\cdots +\rho_{m_k}^2
      \over \rho_i^2\cdots +\rho_{m_k}^2\right)}^{m_k-i}
\end{equation}
They describe how shapeful the perturbations are. 
If the perturbations are stringy then
the maximal moduli will be much larger than the minimal moduli.
If the perturbations are circular, then all moduli will be roughly
the same. This favours some combinations of angles, which are
otherwise uniformly distributed. In general any shapeful picture
defines a line on the ring spectrum $\theta_i$.
A non-Gaussian theory ought to define a set of probable smooth
ring spectra peaking along a ridge of typical shapes.

We can now construct an invariant for each adjacent pair of
rings, solely out of the moduli. If we order the $\rho_i$ for each
ring, we can identify the maximum moduli. Each of these moduli
will have a specific direction in Fourier space; let 
 ${\bf k}_{max}$
and ${\bf k}^{'}_{max}$ be the directions where the maximal moduli
 are achieved.
The angle
\begin{equation}
  \psi(k,k')={1\over \pi}{\rm ang}({\bf k}_{max},{\bf k}^{'}_{max})
\end{equation}
will then produce an inter-ring correlator for the moduli, the
{\it inter-ring spectra}. This 
is uniformly distributed in Gaussian theories in $(-1,1)$. It gives
us information on how connected the distribution of power is between
the different scales. 

We have therefore defined a transformation from the original modes
into a set of variables $\{r,\theta,\phi,\psi\}$. The non-Gaussian
spectra thus defined have a  particularly simple distribution
for Gaussian theories. They also comply
with the uniformity requirement we have place on non-Gaussian spectra
in the discussion at the start of this Section.
We shall call perturbations for which the phases are not uniformly
distributed localized perturbations. This is because if perturbations
are made up of lumps statistically distributed but with well defined 
positions then the phases will appear highly correlated. We shall
call perturbations for which the ring spectra are not
uniformly distributed shapeful perturbations. We will identify later
the combinations of angles which measure stringy or spherical shape of the
perturbations. This distinction is interesting as it is in principle
possible for fluctuations to be localized but shapeless, or more 
surprisingly, to be shapeful but not localized. Finally we shall call 
perturbations for which the inter-ring spectra are not uniformly
distributed, connected perturbations. This turns out to be one of
the key features of stringy perturbations. These three definitions
allow us to consider structure in various layers. White noise
is the most structureless type of perturbation. Gaussian fluctuations
allow for modulation, that is a non trivial power spectrum $C(k)$,
but their structure stops there.
Shape, localization, and connectedness constitute the three next
levels of structure one might add on. Standard visual structure
is contained within these definitions, but they allow for more
abstract levels of structure. We will show in 
Appendix II what these concepts mean with reference to visual
structure.

In the formulation above there is a minor flaw which we found
inconsequent, given the practical advantages gained. This flaw
is spelled out and corrected in Appendix I, but we have
chosen not to do so in the main body of this paper. In Appendix I
we also mention what can be done with the phases $\phi$.
This is however outside the scope of this paper, where
we have decided to investigate the practical applications
of the less investigated ring and inter-ring spectra.

\section{Applications}
Historically much attention has been payed to the non-Gaussianity in
the phases $\phi$. As mentioned above, it has frequently been assumed
that the prescription of random phases in Fourier space leads to
Gaussian perturbations. Evidence of peculiar behaviour
of the phases was shown in numerical simulations of CMB anisotropies
from cosmic strings \cite{BBS,Pom}. Little attention has been given 
to the $\rho$'s. In the following three applications we will focus on the
statistics only involving the $\rho$'s and show that, in these cases,
 they are good non-Gaussian indicators. 

In all these examples we will consider maps with $160^2$-pixels with
no noise; it has been customary to apply the various standard statistics
to the raw non-Gaussian signal superposed with small scale Gaussian noise,
but no attempt has been made at studying the effects of large scale
Gaussian fluctuations. As we will argue there are physically motivated
reasons for doing so. With the intent of keeping the different effects
separate we will analyse this latter case. The addition of noise
should be studied when considering specific observing strategies. 

We will quote all values of the wavenumber, $k$, using uncorrelated
mesh units. I.e. following the discussion of Section II, we will start
labelling the wavenumbers in unit intervals, from the smallest up to
the largest. The width of the rings are therefore $\Delta k=1$.

\subsection{Unresolved point source on a Gaussian signal}
As a first application of these statistics let us consider a Gaussian
signal when non-Gaussian foregrounds are present. We know that this 
is the case in real CMB measurements and there exist a series of
techniques which allow one to separate the two signals using
a combination of spectral and spatial information. A more difficult
situation occurs when one considers unresolved point sources. In this
case, either one uses additional information about the patch of the
sky one is observing \cite{catpeople} or one has to make 
assumptions and the best one can achieve is to subtract them on a 
statistical basis.

Let us consider a simple case which illustrates the weakness of
current methods for checking non-Gaussianity but highlights the
strengths of our technique. 
Suppose that the field is sufficiently small for only a small number
of point sources to be present. Also suppose that the signal 
is Gaussian and that it has
reached the Silk damping tail. The probed spectrum will then
go to white-noise at the scale of the field size, but converge
to the  raw spectrum otherwise \cite{hobsmag}. A
fitting formula for the power spectrum of the Gaussian
signal is 
\begin{equation}\label{exppower}
  P_g(k)=\alpha\exp{\left( -k^2\over 2k_g^2\right)}
\end{equation}

On top of this one must 
either firmly believe that the signal is Gaussian, or that the
signal is non-Gaussian, but of a distinctively different shape.
Now let a single unresolved source be present in the field. Let
the source be perfectly circular, and have a Fourier space falloff
of the form
\begin{equation}
  P_{ng}(k)={1\over 1+(k/k_c)^4}
\end{equation}
The phases are all correlated and arranged so as to center the configuration
and the angles $ \theta$ correspond to a perfectly circular
configuration. All moduli are exactly equal the square root 
of the power spectrum. This is a shapeful, localized, and connected
perturbation, visually recognizable as highly non-Gaussian (see 
Fig. 2. 
Although we are using it as a toy model for an unresolved source, 
this is inspired by a spot produced by a texture undergoing
perfect spherically symmetric collapse.

In Fig. 2. we show the point source, and the signal mixed
with the point source for the case
$\alpha=3$, $k_c=0.1$, and $k_g=5$. What has started as visually
very non-Gaussian disappears completely with the addition of
Gaussian signal. A real space subtraction of the source is bound
to fail.
From inspecting the histogram of temperatures at each realization
one finds that, comparing with a purely Gaussian map with the same
overall power spectrum, they look the same (see Fig. 3).
A more thorough analysis would lead us to calculate the skewness, $\alpha_3$,
and kurtosis, $\alpha_4$, of the maps,
\begin{eqnarray}
\alpha_3&=&C^3(0,0)/(C^2(0))^{3/2} \nonumber \\
\alpha_4&=&(C^4(0,0,0)/(C^2(0))^2)-3
\end{eqnarray}
 or better yet, estimate the distribution of $\alpha_3$ and
$\alpha_4$. In Fig. 4 we superpose histograms of
of skewness (left panel) and kurtosis (right panel) for the
non-Gaussian theory and for the purely Gaussian theory; clearly the
Gaussian behaviour on large scales is dominating the effect of the point
source.

One useful statistic to apply is the accumulated density of peaks
above a given threshold. It was shown in \cite{BE} that, for a Gaussian
field, the density of peaks over a threshold $\mu\sigma$ where
$\sigma=\sqrt{\langle|{{\delta T}\over T}|^2\rangle}$ is approximately given
by:
\begin{eqnarray}
N_{peaks}(\mu)={{1}\over{4\pi\sqrt{3}\theta_*^2}}{\rm max}[1,({6 \over
\pi})^{1/2}\gamma^2\mu \exp(-\mu^2/2)+{\rm erfc}\{{{\mu}\over
{[2(1-2\gamma^2/3)]^{1/2}}}\}]
\end{eqnarray}
where $\gamma$ and $\theta_*$ are dimensionless ratios of the first
three moments of the random field. We can apply this statistic to
our maps, and in Fig. 5 we compare the peak density of
the non-Gaussian maps with that of the pure Gaussian theory (with
the same power spectrum). Although there is a slight difference
for low (negative thresholds) the two peak densities are 
essentially indistinguishable.

We can now apply the approach we have devised. The non-Gaussianity 
will only become evident on small scales, i.e.  for large $k$s in
the Fourier plane. In fact we can find an analytical expression for
 the ring spectrum of a perfectly circular configuration: all
moduli are equal to the same value $\rho_i=I$. Then the ring spectrum is
\begin{equation}\label{thetacirc}
  {\theta}_i^{\rm circ}={\left(m_k-i\over m_k-i+1\right)}^{m_k-i}
\end{equation}
For large values of $m_k$ this ring spectrum is approximately
$1/e$ for all $i$, until $i$ approaches $m_k-1$, where the spectrum rises
to $1/2$. 
As shown in Fig 6 (left panel) the ring spectrum
at a low $k$ is indeed consistent with a uniform distribution (the
$\theta_i$s are uniformly distributed between 0 and 1). 
As $k$ increases the angles $\theta_i$ start accumulating 
around the circular ridge. Soon the point-source dominates the signal,
a fact evidenced by a perfectly circular ring spectrum. Well
into the non-Gaussian region of Fourier space (where the Gaussian
signal is strongly suppressed) we find a clean signal as shown
in Fig 6 (right panel).

This example illustrates the main idea and the main weakness behind our
technique. The main idea consists of trying to identify the 
particular scale on which
non-Gaussianity is evident and clearly this is best done in Fourier
space. In this case (with no experimental small scale noise) one
simply needs to look at $k$s on sufficiently small scales; the
inclusion of Gaussian noise would introduce and outer limit in
Fourier space, reducing the region of non-Gaussianity to a finite ring.

As for the main weakness we point out that the
shape spectrum, $\theta_i$, is sensitive only to the global
shape of the map. While one point source leads to a very clean
distribution of power around rings in Fourier space, if one has
more than a few point sources then this will become less clear. Although for
a set of N sources one will have a very distinct signal (a smooth
line as opposed to a random distribution of $\theta_i$) it becomes
more difficult to distinguish the sources on a firm basis 
from a purely Gaussian
signal. This leads us to establish the best operational strategy
for this method to work: choose small
fields and analyse them separately. In doing this one will be probing
the scales on which non-Gaussianity becomes dominant with less
objects to pick out. The fact that interferometric measurements
of the CMB are constrained to small fields leads us to believe
this to be a sensible prescription for $uv$plane data analysis.
Recent experience with such measurements \cite{catpeople} seems to
indicate that indeed in each field there are only a few problematic
sources (maybe one or two).

\subsection{A cosmic string with a Gaussian background}

One of  the best motivated theories of non-Gaussian structure
formation is that of cosmic strings. Following a primordial
phase transition, line-like concentrations of energy could form in
certain grand unified theories \cite{VS}. This network of
strings would then evolve into a self-similar scaling regime,
 perturbing matter and radiation during its evolution. The non-linear
evolution of the strings should lead to a non-Gaussian distribution
of fluctuations; more specifically, the effect of
strings on radiation after recombination should lead to very distinct 
line-like discontinuities in the CMB\cite{ks,BBS}, the Kaiser-Stebbins
effect. In \cite{BBS} the authors solved Einsteins
equations sourced by a high-resolution simulation of an evolving
string network. They argued that the non-Gaussianity was due
to non-random phases and illustrated this by generating maps with
the same amplitudes but randomized phases and comparing the two.
A battery of tests have been used to quantify these non-Gaussian
features, in some cases with the inclusion of instrument noise and
finite resolution: in \cite{gott} the authors looked at gradient
histograms and the statistics of the genus of excursion sets, in
\cite{periv} an analytical fit to the kurtosis of a string map was
proposed and in \cite{Pom} a multifractal analysis of one
dimensional
scans was proposed.

More recent studies of the evolution of string perturbations in
the CMB indicate that the Kaiser-Stebbins effect is obscured on
subdegree scales by fluctuations generated before recombination
\cite{ACFM},
and that these perturbations look very Gaussian \cite{turok}. None of
the previous statistical tests have taken this into account.
A careful analysis of the behaviour of these two contributions, however,
indicates that the non-Gaussian features may become
dominant again on very  small scales: perturbations seeded
before recombination will be exponentially suppressed by Silk damping
 \cite{silk} on small scales, while the Kaiser-Stebbins effect
will lead to a $k^{-2}$ behaviour. This is an ideal situation
for using our statistic. We can evaluate the non-Gaussian spectrum
on scales where the non-Gaussian signal is expected to dominate,
and see if it shows any evidence for
deviation from the background Gaussian distribution.

If we consider the case of a very small field, we expect to have at
most one segment of string crossing the patch. This would be the case 
for a field of a fraction of a degree. It is instructive to consider
the case of a smooth, straight string. Here the signal is maximally 
non-circular and all of the power in the ring is concentrated on one 
of the modes $\rho_s=m_kI$, with $\rho_i=0$ for $i\neq s$. 
For such a configuration the ring spectrum is 
\begin{eqnarray}
  {\theta}_i&=&1{\rm \quad for} \quad i<s\nonumber\\
  {\theta}_i&=&0{\rm \quad for} \quad i=s\nonumber\\
  {\theta}_i&=&{0\over 0}{\rm \quad for} \quad i>s\nonumber\\
\end{eqnarray}
The last angles are undefined in the same way that the angle $\phi$
in the normal 3D polar coordinates is undefined for points along the $z$
axis. The point remains that the configuration corresponds to a single
point on the $m_k-1$ dimensional sphere, and that therefore has probability
zero in a Gaussian theory. For display purposes one may then also
fix the remaining angles at some particular but arbitrary value.
We define $0/0=0$. 

For a perfectly straight string non-Gaussianity
is so extreme that it is will be visually evident even with a very large
amount of background Gaussian noise.
The situation changes dramatically, however, for 
the more realistic case when the string is rough, or structured.
This is the picture that emerges from high resolution numerical 
simulations \cite{hires}. The intercommutation of strings will build up kinks
and cusps along a string which will only stabilize once gravitational
radiation becomes important. Again most of the power will be
concentrated along one or a few modes, leading to a well defined
spectrum up to some maximum $i$. For larger $i$'s the spectrum
will be close to $0$ or ill defined in the same way as the straight
string case.

Having played with a string code, 
we have chosen to model the string as a directed
Brownian walk along the patch we are considering. 
We then modelled the effect of the Gaussian background on these
scales in the same way as in the previous example. We superimpose
a background Gaussian signal with the power spectrum given in \ref{exppower}.
In Fig 7 a 
we show an example of a $(160)^2$-pixel map ($20$ arcmin$^2$) of the  non-Gaussian signal and 
in Fig 7 b  we superimpose a Gaussian background with a $k_g=26$
and with 5 times the overall amplitude of the non-Gaussian signal.
Clearly the beautiful Kaiser-Stebbins effect is now beyond
what we can recognize visually. One must therefore resort to
more abstract tests.

We first applied to our maps some of the standard tests. It has been argued 
that the skewness and kurtosis of the gradient of the temperature
anisotropy field should be a good indicator of string
non-Gaussianity. Skewness should be very sharply peaked at $0$,
(the patterns caused by the string are very symmetrical in terms of
amplitude), and kurtosis should be larger than the Gaussian
\cite{periv}. In Fig 8 we show histograms of skewness and
kurtosis made from an ensemble of 400 realizations. Clearly the 
string with a Gaussian background is indistinguishable from the
purely Gaussian sky.

A more elegant statistic involves working out the Euler characteristic
of the maps, given a threshold. The procedure is straightforward:
given a threshold $\mu\sigma$ one evaluates
the difference between the number of isolated hot regions and cold
regions
with regard to $t$. For a Gaussian field the mean genus 
is
\begin{eqnarray}
\Gamma\propto \mu e^{-{{\mu^2}\over 2}}
\end{eqnarray}
It was argued in \cite{gott} that this would be a 
good indicator of non-Gaussianity
for strings. In Fig 9 we show the Euler characteristic
averaged over 100 runs for the string with a Gaussian background and
for a purely Gaussian map with the same power spectrum. Again we find
no significant difference between the two.

Finally we have applied to these maps our technique.
We first looked for the  distribution of the $\theta_i$s in 
rings where the non-Gaussianity is evident. Due to the random nature
of the structure on the string, the signal in the ring spectrum
won't be as cleanly defined as for the straight string case. 
We therefore looked at a large number of maps in order to plot
$\theta_i$s with cosmic/sample invariance errobars.
For plotting purposes we shall give error bars as regions
of probability larger than $1/e$. This corresponds to a 
1-$\sigma$ errorbar if the distribution is Gaussian, but generalizes
the concept of a 1-$\sigma$ errobar to more general distributions.
In particular the concept may be applied to a uniform distribution,
which does not even have a peak.
In Fig 10, the shaded region is  where the $\theta_i$s have
more than $1\over e$  probability of being; the ring has $k=70-75$
(for a $160^2$-pixel map) and we clearly see a ridge towards
 the left hand side. For rings at low $k$ this ridge blurs
into the standard Gaussian prediction.

A more striking statistic is the inter-ring spectrum. 
In Fig 11, we have shaded the region where $\psi$s have more
than $1\over e$  probability of being. It is clear that for 
low values of $k$ the Gaussian background dominates, and the
various rings are essentially uncorrelated. However
above a certain threshold, subsequent modes are tightly
correlated. As argued above, most of the power is concentrated
along one direction of each ring. What we see here is that this
direction is strongly correlated between rings.
This quality we labelled as connectedness. We see that
strings' connectedness is a robust non-Gaussian feature,
even when all else seems to fail.

\subsection{Evasive non-Gaussian theories}
We finally present a strongly non-Gaussian theory on all scales 
which evades detection by several traditional 
non-Gaussianity tests. 
Consider a theory with a power spectrum as in (\ref{exppower}), 
say with $k_g=10$, in uncorrelated mesh units. 
Let the phases $\phi$ and inter-ring
correlator angles $\psi$ be uniformly distributed. However
let the ring spectra ${\theta}(k)$ for all rings $k$
be the circular ring spectrum ${\theta}^{\rm cir}(k)$
(cf. Eqn.~\ref{thetacirc}) with infinite probability density. 
Thus we have theory
of delocalized, disconnected spheres. In Fig. 12 we
show a realization of this theory (call it theory $T_1$) and also
a Gaussian realization, that is, a realization of a theory (call it
$T_2$) which differs only in that the ${\theta}(k)$ are now
uniformly distributed.

Theory $T_1$ is strongly non-Gaussian. The set of all of its realizations
has measure zero in any Gaussian theory. In other words the cosmic
confusion between the two theories is zero, where cosmic confusion
is defined as the percentage of common skies generated by the two 
theories \cite{mag1}. 
If $Q$ is the set of all map variables, and if $F_1(Q)$
and $F_2(Q)$ are their distribution functions in the theories $T_1$
and $T_2$, then the cosmic confusion between the two theories is
\cite{mag1}
\begin{equation}
  {\cal C}(T_1,T_2)=\int dQ\, \min{(F_1,F_2)}
\end{equation}
In terms of the variables $Q=\{C(k),{\theta}(k),\phi,\psi\}$
we have 
\begin{eqnarray}
  F_1&=&\prod_k\chi^2_{N_k}(C(k))\prod_{\phi}\frac{1}{2\pi}
             \prod_{\psi}\frac{1}{2\pi}\prod_{{\theta}}
             \delta({\theta}-{\theta}^{\rm circ})\\
  F_2&=&\prod_k\chi^2_{N_k}(C(k))\prod_{\phi}\frac{1}{2\pi}
             \prod_{\psi}\frac{1}{2\pi}
\end{eqnarray}
so that ${\cal C}(T_1,T_2)=0$.

Although we have as yet no physical
motivation for such a theory, we believe it to be a good example where
the traditional beliefs about non-Gaussianity do not hold; in spite of 
its strong non-Gaussianity this theory evades all
tests we have applied to it. Visually the maps 
produced by the theory look very Gaussian. We can apply all
the test we have introduced in the previous two sections
with rather spectacular failure.
Plotting temperature histograms reveals
a very Gaussian distribution (see Fig. 13). One may
convert these histograms into moments, with the same result. 
The Sections of the $n$-point function which may be computed in 
practice are also very Gaussian.
In Fig. 14 we have plotted the average and 1-sigma errobars
for the collapsed 3-point correlation function for $T_1$ and $T_2$
as inferred from 100 realizations. In Fig. 15 we plot
histograms of kurtosis for the two theories. Clearly they are not
good discriminators between the two theories.
 We can estimate the number of peaks over a given threshold for the
two theories. In Fig. 15 we plot the total number of peaks
above a given threshold for $T_1$ and $T_2$. In Fig. 16 we
find the Euler characteristic for the two theories. Once again they
are indistinguishable.

Nevertheless all rings of the $uv$-plane show a ring spectrum
which is perfectly circular, without any variance. Any sky, and any $k$,
produces a ring spectrum as the one in Fig. 18,
obtained from the same realization used above, for the ring $k=11$.
\section{Discussion}

In this paper we have proposed a transformation of variables in
Fourier space which produces non-Gaussian spectra with a particularly
simple probability distribution 
function for a Gaussian random field. We have focused on a subset
of these, the ring spectra, $\theta_i$, and the inter-ring
spectrum, $\psi$, which contain information about the moduli of the
Fourier modes. We have presented a few examples where they are
good qualifiers of non-Gaussianity.

A number of comments are in order with regards to the limitations
of these statistics. To begin with these statistics
are tailored for data in Fourier space. To actually apply these
statistics to real space data will involve
non-local transformations which may complicate the procedure. However,
in the examples which we have worked out, the non-Gaussianity becomes
apparent on small scales. Therefore one is forced to consider experiments
with the best possible resolution. These are interferometric devices
where the data is measured directly in Fourier space. Another possible
shortcoming of these statistics is that they are sensitive to the
global shape of the data set or map. This means that if one has
many non Gaussian features (such as many point sources or many
segments of string) then both the ring spectrum and the inter-ring
spectrum will look more Gaussian. This can only be avoided by looking at
small fields. But once again this is the situation
favoured by interferometers. One is limited to small fields (although
one can mosaic over reasonably large patches of sky, \cite{CWF}) and
experience in \cite{catpeople} indicates that very few unresolved
sources will be present.
In a interferometric  search for string segments, one would restrict oneself
to fields of less than ${.5^{\rm 0}}^2$ and still have 
a $90\%$ probability of actually seeing a string, but not more than one.

We have not included the effect of small scale noise in the examples
we considered. In those cases the signal was already sufficiently
corrupted for it to be difficult to identify the non-Gaussian
features. In fact, what one finds is that large scale Gaussianity seems
to be more devastating (in terms of erasing non-Gaussian features)
than small scale, noise-related, 
Gaussianity. Clearly one has to include the two
effects if one wants to apply these techniques to data but the
details are  dependent  on each experiment. The statistics defined are
non-linear statistics in the data which means care must be had when
considering the effect of noise. A case by case analysis of the
different observing strategies will have to be made.  Again, the
fact that the small scale noise in interferometers 
increases as a power-law with scale,
as opposed to exponentially as in the case of a single dish
experiment, indicates that interferometric devices are the best
instruments for testing for non Gaussian features.
One immediate goal will be to design the ideal experiment for
detecting the Kaiser-Stebbins effect. This should include a
careful analysis of theoretical uncertainties (such as the
amplitude of fluctuations at last scattering) as well as
the real life complications mentioned above.

We have focused on statistics with the moduli, $\rho$, and have 
not developed in any detail, or applied to any example, statistics
with the phases, $\phi$. It is conceivable that much information
can be extracted from their behaviour. In fact, a generic
feature of physically motivated non-Gaussian models is localization,
which, as we have argued is governed by the phases. Although we have
organized the information that can be extracted from a finite data
set in systematic way, it is important to define a useful set of
statistics in terms of the phases. We will do so in \cite{fermagagain}.

ACKNOWLEDGMENTS: We thank A. Albrecht, S. Hanany, J. Levin, J. Silk
and L. Tenorio
for interesting conversations.
P.F. was supported by  the
Center for Particle Astrophysics, a NSF Science and
Technology Center at UC Berkeley, under Cooperative
Agreement No. AST 9120005.
J.M. thanks St.John's College, Cambridge, and the Royal
Society for support, and also Joe Silk and CfPA for
hospitality.

\pagebreak
\pagestyle{empty}
\section*{APPENDIX 1: Invariant shape and phase spectra}
The fact that the uncorrelated mesh points are somewhat undefined
makes the search for invariant quantities a pedantic matter.
For this reason we decided to define shape variables $\theta$
which strictly speaking are not invariant. The inter-ring spectrum, 
on the other hand, is already invariant.

It is however possible  to define invariant shapes but they are  
more complicated. Under a rotation the moduli $\{\rho_i\}$ suffer a 
cyclic permutation. Hence the 2D-rotation group has now become 
discrete and so it will not discount a degree of freedom. Nevertheless the 
angles $\{\theta_i\}$ defined from them will not be invariant
under rotations (translations do not affect the $\{\rho_i\}$).
A way around this is to order the $\{\rho_i\}$ so that the last
$\rho_i$ is the largest. The angles $\{\theta_i\}$ produced from
the ordered $\{\rho_i\}$ will then be properly invariant.
They will also always be defined.
The joint distribution of the ordered $\{\rho_i\}$ is proportional
to the joint distribution of the unordered ones. In fact the Jacobian
of any variable interchange is 1. One may at most pick a proportionality 
constant from adding over all the branches of the transformation.
Hence the whole argument in Section III still applies, and the new, ordered, 
$\{\theta_i\}$ will still have a joint distribution which is
uniform. However the new $\{\rho_i\}$ and $\{\theta_i\}$ are now
dependent random variables, not because their joint distribution
does not factorize, but because the domain of some of the variables
depends on the others. This results from  $\rho_i\leq\rho_m$.
This has several unpleasant consequences. For instance the 
marginal distribution of any of the $\theta$ is now not the
factor appearing in the joint distribution function. Hence
the marginal distribution of the properly invariant  $\theta$'s
is not uniform, although their joint distribution is.
All in all we found the $\theta$'s we have defined in the main
body of this paper more practical to use, as they are much 
better behaved in Gaussian theories.

The phases $\phi$ defined in the main body of this paper are also not
invariant. Under a rotation they suffer a cyclic permutation,
whereas under a translation by a vector ${\bf t}$ they
transform as $\phi({\bf k})\rightarrow \phi({\bf k}) +{\bf k}
\cdot {\bf t}$. The phases $\phi({\bf k})$ may be seen as an 
antisymmetric real scalar field on the space ${\bf k}$. 
In this language the field gets rotated under a (real space) rotation,
and acquires a dipole under a (real space) translation. One can build
invariants out of the phases, therefore, simply by subtracting
the dipolar component of the field, and averaging over angle.
This can be done in many different ways, to be explored more
thoroughly in a future publication. Here we simply outline
one possible strategy. Let us in each $\Delta k=1$ ring
apply an angular Fourier transform to the phases:
\begin{equation}
  \phi({\bf k})=\phi(k,\beta_k)={\sum_m}\phi(k,m)e^{im\beta_k}
  ={\sum_m}\phi_c(k,m)\cos(m\beta_k)+\phi_s(k,m)\sin(m\beta_k)
\end{equation}
Then under a translation the $m=1$ mode transforms as
\begin{eqnarray}
  \phi_c(k,1)\rightarrow \phi_c(k,1) + kt\cos\beta_t\\
  \phi_s(k,1)\rightarrow \phi_s(k,1)+ kt\sin\beta_t
\end{eqnarray}
whereas all other modes are invariant. One may then simply
throw away the $m=1$ mode, the other ones making up a 
localization ring spectrum. The distribution of these
in Gaussian theories is again not simple, and we shall look
for something better than this. This procedure however
does have the advantage of reacting to individual shapes
an localization properties rather than global ones.

For any pair of adjacent $\Delta k=1$ rings we have subtracted
two modes too many. These should be returned in the form of two
inter-ring phase invariants, such as
\begin{eqnarray}
 \Phi_c(k) &=&{\phi_c(k+1,1)\over k+1}-{\phi_c(k,1)\over k}\\
 \Phi_s(k) &=&{\phi_s(k+1,1)\over k+1}-{\phi_s(k,1)\over k}
\end{eqnarray}
Again this is but one example of a possible invariant
made out of phase gradients, to be explored better in
our future work.

\section*{APPENDIX 2: Visual interpretation of non-Gaussian
spectra}
The decomposition $\{C(k),{\theta},\phi,\psi\}$ has 
an immediate physical interpretation. The angles $\theta$
reflect the angular distribution of power, and therefore reflect
shape. The phases $\phi$ transform under translations and so contain
the information on position and localization of the structures in the field.
The angles $\psi$ correlate different scales, and therefore tell us
how connected the structures are. For a Gaussian random field
the variables $\{{\theta},\phi,\psi\}$ are all uniformly
distributed reflecting complete lack of structure besides the power spectrum.
In terms of the various levels of structure considered we
can then characterize Gaussian fluctuations as shapeless, delocalized
and disconnected. By comparison with a Gaussian we may then define
structure at different levels. We will say that fluctuations for which
${\theta}$ are not uniformly distributed are shapeful.
If the $\phi$ are not uniformly distributed we shall say the
fluctuations are localized. If the $\psi$ are not uniformly distributed 
the fluctuations are connected. Although visual structure has
room within these definitions, they are considerably
more abstract and general. We may consider highly non visual
types of structure such as shapeful but delocalized fluctuations
or disconnected localized stringy fluctuations. In this sense
we regard our formalism as a robust definition of structure, which goes 
beyond what is visually recognazible and so is tied down to 
our particular and narrow path of natural selection. We may
imagine an alien civilization with Fourier space eyes (say
interferometric eyes \cite{nota}), and a brain trained to recognize
Fourier space structure at many different levels, 
structure that would seem totally non obvious
to our human eyes.

To illustrate the limitations of human vision
we shall now destroy highly structured 
maps level by level, that is Gaussianize only one of the
variable types $\{{\theta},\phi,\psi\}$. Initially there
will be structure at every level, shape, position, and connectedness.
We will remove structure gradually, a fact not disasterous
for the alien civilization referred above, but which will
illustrate the  limitations of the human visual method for recognizing
non-Gaussianity. In Figure 19 we play this
game with a sphere. We depict a spherical hot spot in real space, then
a shapeless sphere, a delocalized sphere, and a disconnected 
sphere.  For the case of a sphere we find that what we recognize as shape
is mostly localization. A shapeless sphere keeps its 
recognizable features. On the other hand a delocalized sphere
loses it characteristic features. Indeed the idea of a shapeful but
non-localized object sounds somewhat surreal for all we can 
visually conceptualize. Nevertheless our formalism
will acuse the strong but not obvious non-Gaussianity exhibited
by a delocalized sphere. 

In Figure 20 we repeat the same
exercise for a map displaying the Kaiser-Stebbins effect from
cosmic strings. Shapeless strings, delocalized strings, and disconnected
strings are shown. Considerable disarray is introduced
in every case, but one may say that disconnected strings as well as
delocalized strings are 
perhaps the most messy of them. This is consistent with the strong signal
in $\psi$ we have found for the case of the realistic Kaiser Stebbins effect.
On the other hand the fact that line-like discontinuities are present 
even for shapeless strings shows how much more structure
there is in the map on top of the structure
which we can recognize. This is important since the beautiful patchwork
is very fragile to the hard realities of noise
and  supperposed Gaussian signal.
In the real world, it turns out, the non-visual feature which
is the connectedness of strings happens to survive much better than the
patchwork (which reflects mostly localization).

\eject
\section*{Captions}

Fig 1: The most general second order invariant (left) is the angular power
spectrum, obtained by multiplying the mode ${\bf k}_1$ amplitude
with the mode ${\bf k}_2=-{\bf k}_1$ amplitude, and averaging over directions.
The result can only depend on $k$. On the right we show one possible 
configuration giving a third order invariant, the one where all 3 vectors
have the same moduli. Then for each $k$ all 3 vectors are determined
from the requirement that they must add up to zero. Averaging
over directions produces an invariant.

Fig 2: The non-Gaussian signal (top left)  and the 
full signal, with the Gaussian superposition (top right)
for $\alpha=3$, $k_c=0.1$, and $k_g=5$.
In the map on the left the skewness is 2.9 and the kurtosis 11.4.
In the map on the right the skewness is -.02  and the kurtosis -.38.

Fig 3: Histograms of the temperature distributions for the non
Gaussian map (solid line) and purely Gaussian map with the same power 
spectrum (dashed line).

Fig 4: Histograms of the skewness (left panel) and kurtosis (right
panel) for the non
Gaussian map (solid line) and purely Gaussian map with the same power 
spectrum (dashed line).

Fig 5: The density of peaks above a threshold,
${\delta T \over T}$, for the non Gaussian theory (solid line)
and the purely Gaussian theory (dashed line). The curves are averaged over
20 runs.

Fig 6: The ring spectra for two rings $k=20$ and $k=50$.

Fig 7: The Kaiser-Stebbins effect  (a- left)  and the 
full signal, with the Gaussian superposition (b-top right)
for $\alpha=5$,  and $k_g=26$.

Fig 8: Histograms of skewness (a-left) and kurtosis (b-right)
of the gradient of temperature anisotropies 
from 400 realizations of a string maps with Gaussian background
(solid lines) and Gaussian realizations with the same power
spectrum (dashed lines).

Fig 9: The mean Euler characteristic, $\Gamma$ as a function of
threshold for a string map with Gaussian noise (solid line)
and for a pure Gaussian map with the same power spectrum
(the shaded region is the 1 $\sigma$ region around the Gaussian mean,
estimated from a 100 realizations).

Fig 10: The ring spectrum for $k=70$ and for $\alpha=5$,  and $k_g=26$.
 The shaded region
represents a probablilty larger than $1 \over e$ for the
the values of $\theta_i$ to occur.

Fig 11: The inter-ring spectrum with for $\alpha=5$,  and $k_g=26$.
 The shaded region
represents a probablilty larger than $1 \over e$ for the
the values of $\psi_i$ to occur.

Fig 12: Realizations of theory $T_1$ (left) and $T_2$ (right).
Theory $T_1$ is a theory of disconnected delocalized perfect
spheres, with zero cosmic confusion with theory $T_2$, which had the
same power spectrum, but is Gaussian.

Fig 13: Temperature histograms for the two maps shown above
(that is non-Gaussian (left) and Gaussian (right)).
The skewness is respectively 0.043 and -0.068, and the kurtosis
-0.042 and 0.229.

Fig 14: The collapsed 3-point correlation function for 
theories $T_1$ and $T_2$. The averages and errorbars were
inferred from 100 realizations in both cases.

Fig 15: The histograms of kurtosis for theory T1 (solid)
and theory T2 (dashed) taken from an ensemble of 1000 realizations.

Fig 16: The density of peaks above a threshold,
${\delta T \over T}$, for theory T1 (solid line)
and theory T2 (dashed line). The curves are averaged over
20 runs.

Fig 17: The average Euler characteristic of theory T1
(solid) and T2 (dashed) averaged over 100 realizations.

Fig 18: All rings  of theory $T_1$ for all realizations show
a perfect circular ring spectrum. Here we show the ring $k=11$.

Fig 19: A spherical hot spot which has been deconstructed at
different levels. On the top left hand panel we have the pure 
non-Gaussian
signal. The angles $\theta_i$ have been
redrawn uniformly on the top right picture. On the bottom left
the phases $\phi_i$ were redrawn unformly. On the bottom right
we applied an independent unformly distributed  rotation on 
all rings in Fourier space.  From top to
bottom and left to right, a plain regular sphere, a shapeless
sphere, a delocalized sphere, and a disconnected sphere.

Fig 20: The Kaiser-Stebbins effect (top left) and its various
stages of deconstruction. The angles $\theta_i$ have been
redrawn uniformly on the top right picture. On the bottom left
the phases $\phi_i$ were redrawn unformly. On the bottom right
we applied an independent unformly distributed  rotation on 
all rings in Fourier space.  Respectively we have strings,
shapeless strings, unlocalized strings and disconnected strings.

\begin{thebibliography}{99}
\bibitem{Peebles2}P.J.E. Peebles, {\it Principles of Physical
Cosmology}, Princeton University Press (1993).
\bibitem{inflCMBR} P. Steinhardt, {\it Cosmology at the
Crossroads} to appear in the {\it Proceedings of  the Snowmass Workshop on
Particle Astrophysics and Cosmology}, E. Kolb and R.Peccei,
eds. (1995)   astro-ph/9502024.
\bibitem{kogut} Kogut {\it et al}, {\tt astro-ph 9601062}
\bibitem{lascampanas}S. Landy et al, {\it Astrophys.J} {\bf 456} L1-L4 (1996).
\bibitem{VS} A. Vilenkin and P. Shellard, {\it Cosmic
Strings and other Topological Defects.} Cambridge University Press,
Cambridge (1994);T.W.B. Kibble, {\it J. Phys.}, {\bf A9} 1387-1398 (1976).
\bibitem{catpeople} M.Hobson, Proceedings of the Moriond conference on CMB,
(1996).
\bibitem{BE} J. R. Bond and G. Efstathiou {\it MNRAS} {\bf 226}
655-687 (1987)
\bibitem{Jusk}N. Vittorio and R. Juskiewicz {\it Astrophys. J.}
{\bf 314} L29-L32 (1987)
\bibitem{Pcoles}P. Coles {\it MNRAS} {\bf 234} 509-531 (1988)
\bibitem{gott}J. R. Gott {\it et al} {\it Astrophys. J.} {\bf 352}
1-14 (1990)
\bibitem{gangui}A. Gangui and S. Mollerach {\it Phys. Rev.} {\bf D54}
in press.
\bibitem{scaramelo}R. Scaramella and N. Vittorio, MNRAS 263 (1993) L17.
\bibitem{periv}R. Moessner, L. Perivolaropoulos and R. Brandenberger
{\it Astrophys. J.} {\bf 425} 365-371 (1994) 
\bibitem{coulson}D .Coulson, P. Ferreira, P. Graham and N. Turok,
{\it Nature} {\bf 368}  27-31 (1994).
\bibitem{pst}U. Pen, D. Spergel and N. Turok, {\it Phys. Rev.} {\bf
D49} 692-729 (1994)
\bibitem{Peebles1}P.J.E. Peebles, {\it The Large Scale Structure of the 
Universe}, Princeton University Press (1980).
\bibitem{randfields}R. Adler, {\it The Geometry of Random Fields},
(Wiley, New York, 1981).
\bibitem{mag1}J.C.R Magueijo {\it Phys. Lett.} {\bf B342} 32-39 (1995).
\bibitem{hobsmag}M.Hobson, J.Magueijo, {\it MNRAS}, in press,
astro-ph/9603064.
\bibitem{Review}M. White, D. Scott and J. Silk, 
{\it Annu. Rev. Atron. Astrophys.}
{\bf 32} 319-370 (1994).
\bibitem{efrev}G. Efstathiou, in {\it Physics of the Early Universe},
eds. J. Peacock, A. Heaven and A. Davis 361-470 (1991).
\bibitem{ks}N. Kaiser and A. Stebbins {\it Nature} {\bf 310} 391-393
(1984)
\bibitem{fermagagain}P.G Ferreira and J. Magueijo (in preparation)
\bibitem{Pom}M.P. Pompilio, F.R. Bouchet, G. Murante and A. Provenzale
{\it Astrophys. J.} {\bf 449} 1-8 (1995)
\bibitem{BBS}F.R. Bouchet, D.P. Bennet and A. Stebbins {\it Nature}
{\bf 335} 410-414 (1988)
\bibitem{ACFM}A. Albrecht, D. Coulson, P. Ferreira and J. Magueijo
{\it Phys. Rev. Lett} {\bf 76} 1413-1416 (1996);
J.Magueijo, A.Albrecht, D.Coulson, P.Ferreira,
 {\it Phys. Rev. Lett} {\bf 76} 2617-2619 (1996);
J.Magueijo, A.Albrecht, P.Ferreira, D.Coulson,
{\it Phys. Rev.} {\bf D} in press.
\bibitem{turok} N.Turok {\it astro-ph/9606087}
\bibitem{silk} J.Silk {\it Astrophys. J.} {\bf 151} 459 (1968)
\bibitem{hires} D. P. Bennett, in ``Formation and Evolution of 
Cosmic Strings'',
eds. G. Gibbons, S. Hawking and T. Vachaspati, (Cambridge University 
Press, Cambridge.
1990); F. R. Bouchet {\it ibid.}; E. P. S. Shellard and B. Allen {\it 
ibid.};
\bibitem{CWF} J. Carlstrom, M. White and P.G. Ferreira, in preparation.
\bibitem{nota}
After all we have Fourier space ears. We thank Dr.K.Baskerville
for this pertinent remark.
\end{thebibliography}
\end{document}